\documentclass[a4paper,11pt]{article}                         
\usepackage{extsizes} 
\usepackage[english]{babel}
\usepackage{graphicx,color}
\usepackage[utf8]{inputenc}
\usepackage{amsmath,amssymb,amsfonts}
\usepackage{graphicx}
\usepackage[hmarginratio=1:1,top=32mm,columnsep=20pt]{geometry} 
\usepackage{multicol}                                            
\usepackage[small]{caption}         
\usepackage[sc]{mathpazo}         
\usepackage{booktabs}
\usepackage{float}
\graphicspath{{\figs}}      

\title{\vspace{-15mm}{\huge Optimizing Beer Glass Shapes to Minimize \\  Heat Transfer -- New Results} }
	\author{\large\it{Cláudio C. Pellegrini}	\\ 
	\small  Federal Univ. of São João del-Rei \\ \small Depart. of fluid and thermal sciences, São João del-Rei, MG, Brazil.  \\ 
	\small Corresponding author: pelle@ufsj.edu.br
}
\date{}

\begin{document}
\maketitle  

\begin{abstract}
	\noindent
	This paper addresses the  problem of determining  the optimum shape for a beer glass that minimizes the heat transfer while the liquid is consumed, thereby keeping it cold for as long as possible. The  proposed solution avoids the use of insulating materials. The glass is modeled as a  body of revolution generated by a smooth curve,  constructed from a material with negligible thermal resistance, but insulated at the base.  The ordinary differential equation describing the problem is derived from the first law of Thermodynamics  applied to a control volume encompassing  the liquid. This is an inverse optimization problem,  aiming to find the shape of the glass (represented by curve $S$)  that minimizes the heat transfer rate. In contrast, the direct problem  aims to determine the heat transfer rate for a given geometry.  The solution obtained here is analytic, and the resulting function describing the relation between height ans radius of the glass, is in closed form, providing  a family of optimal glass shapes that can be manufactured by conventional methods. Special attention is payed to the dimensions and the capacity of the resulting shapes.
		
\end{abstract}

\section{Introduction}

An effective pedagogical strategy in mathematical and physics education involves demonstrating to students how the theories taught in the classroom can be used to address everyday problems.
Nevertheless, this proves to be a difficult task more often than not. 

Few everyday problems  in physics possess a level of simplicity  that  allows for analytical treatment, while still capturing  all  phenomena involved. Introducing too many simplifications may render the problem  unrealistic\footnote{For example, consider an   airplane  wing represented by a perfectly smooth, infinitely thin  flat plate inclined by an angle $\theta$ in relation  to a  frictionsless airflow.}. If too few simplifications are emploied, the problem may become unsuitable for analytical treatment\footnote{The same airplane wing now is subjected to  turbulent airflow and presents a rough surface defined by given functions $y_{up}(x)$ and $y_{low}(x)$, where $x$ is the position along the chord  and $y_{up}$ and  $y_{low}$ are respectively the upper and lower half thickness at a given value of the position $x$}. 

In the field of mathematics, finding practical  problems that are easy to understand but can be solved by straightforward techniques, is an even more challenging task. 
The four-color theorem\footnote{Given any separation of a plane into contiguous regions, producing a figure called a map, no more than four colors are needed to color the regions of the map, so that no two adjacent regions have the same color.} [Appel and Haken 1977 and Appel et al. 1977] and the hairy ball theorem\footnote{You can't comb a hairy ball without creating a whirlpool.} [Eisenberg et al., 1979] are traditional examples. Both require considerable expertise to  understand the solution. More recently, ideas about how to divide a pizza into an arbitrary number of unconventionally divided parts  Humenberger 2015  have also gained interest, but is equally challenging to understand.

Motivated by the last problem, [Pellegrini 2019\footnote{In Portuguese} and Pellegrini 2024] envisioned an analysis related to the beverage that arguably -- or not -- pairs best with pizza: beer. Using the simplest of the mathematical optimization tools, the first derivative test, the author obtained a function describing the shape of a family of optimized glasses. Nevertheless, most glasses were quite large, with volumes ranging from an impractical two liters to an  outrageous 103 liters.
 
Building on the results of [Pellegrini 2024], this paper revisits the problem of finding the optimum shape for a beer glass, such that the heat transfer rate is minimized, to keep the beverage cold for as long as possible, while it is consumed {\sl with focus} on obtaining glasses of practical proportions and capacity. 


This is an inverse optimization problem,  in the sense that the objective is to find the geometry that minimizes the heat transfer rate as opposed to the direct problem, where the objective would be to find the heat transfer rate associated with a given geometry. The intention is to encourage physics, mathematics and engineering students to develop a rational approach to real problems, abandoning the fairly common concept that "theory in practice is different". As a final contribution, nevertheless, the study also aspires to help improve  the drinkability of our beers.

A brief search on the literature shows literally  hundreds of articles focused on analyzing practical problems across diverse areas of physics, but nothing was found on heat transfer, except for  [Planinsic and Wolmer 2008], which  addresses the heating of cheese. 

The problem of keeping a liquid contained  in a reservoir at the lowest possible temperature may be solved, as will be shown later, by finding a surface that minimizes the area-to-volume ratio of the reservoir. The Greeks  knew that  the answer to the two-dimensional  version of this  problem was the circle, even though they could not prove this rigorously. Later findings showed that the solid possessing the lowest surface-to-volume ratio was the sphere. However, formal proofs for both cases had to wait until the 19th century. In contemporary times, this problem is addressed through the application of the isoperimetric inequality\footnote{Let $\gamma$ be a closed, piecewise differentiable plane curve of class $C^1$. Let $L$ and $A$ be its perimeter and surface, respectively. The isoperimetric inequality establishes that $4\pi A \leq L^2$, with the equality valid if $\gamma$ is a circle.} in three dimensions (for which there is a variety of proofs) or through variational calculus concepts.  
 
Applications of surface-to-volume ratio optimization are numerous and extend to practically all areas. In chemistry, there are studies involving reactions of all types, such as combustion (in engines or fires), drying and humidification of particles. In biology, there are studies involving exchanges through the skin of living beings and the membrane of cells, microorganisms or organelles. In engineering, there are studies of heat transfer in reservoirs and heat and mass transfer in systems subject to phase change. In atmospheric sciences, there are studies involving the formation of raindrops, hail and snowflakes, as well as evaporation in vegetables and bodies of water. In pharmacology, there are studies on the absorption of medications. It is unnecessary to point out the existence of a series of multidisciplinary studies, with an interface between different areas of study. 

A common factor in all the  above mentioned studies, is the fact that the heat exchange surface does not change its shape during the process. This, however, is not the case when considering the heat exchanged by a glass of liquid with the surroundings during consumption. Even in  a cylindrical glass, the total exchange surface undergoes changes in shape during consumption. As the liquid level lowers, the side area reduces while the upper area remains preserved. 

Therefore, the question we answer here is: what is the optimal shape  that minimizes the heat transfer on a glass of liquid being consumed and with feasible proportions? In other words, what is the ideal  beer glass in a (as much as possible\footnote{What exactly does the author mean by ``as much as possible''? Well, the intention here is to find an \textit{analytical solution} to the problem, for reasons that will be  detailed in the Conclusion section. Therefore, ``as much as possible'' really means ``as far as I am able to find an analytical solution to the mathematical model proposed.''}) realistic scenario?

\section{Mathematical model}
The process  is quite straightforward here: a request is made for a beer, the waiter delivers it, it is served,   it is consumed. Repeat. 

Once poured in the glass, the beer begins to exchange heat with its surroundings, a process that lasts until it attains thermal equilibrium with of the environment (including the glass), or  is finished, both result that essentially  nobody wants. Depending on the initial temperature difference between the beer and the surroundings, within a short period of time the drink  may become unsuitable for consumption. In the most favorable  case, the environment  at 10 \textdegree C and the beer\footnote{Who drinks beer below 10 \textdegree C anyway?}   at 4 \textdegree C, personal experience shows that as long  as 30 minutes may pass before the beer gets warm. In the most critical scenario, such as at the beach on a 38 \textdegree C  windy day, as few as 3 minutes  may be sufficient (again based on personal experience, exhaustively repeated) to render the beer undrinkable. 

There are several established practical methods to  decrease  the beer's heat transfer with the environment. The use of insulation tubes made of expanded polystyrene (EPS or Styrofoam${}^{\text \textregistered}$) is probably the most common and it is also used for beer bottles in Brazil. The use of handles on mugs is also a common method to isolate the  consumer's  hand heat from  the drink inside the container. The habit of maintaining a layer of foam on the top of the beer  acts as a thermal insulator due to its low conductivity. In addition, it also prevents excessive loss of CO${_2}$. All those methods have in common the fact that they are  {\sl passive}, i.e., they do not depend on any heat transfer device or substance. That is exactly the approach we shall use here. 

\begin{figure} [htb]
   \centering 
   \includegraphics[trim={50mm 0 50mm 0}, clip,width=0.5\columnwidth]{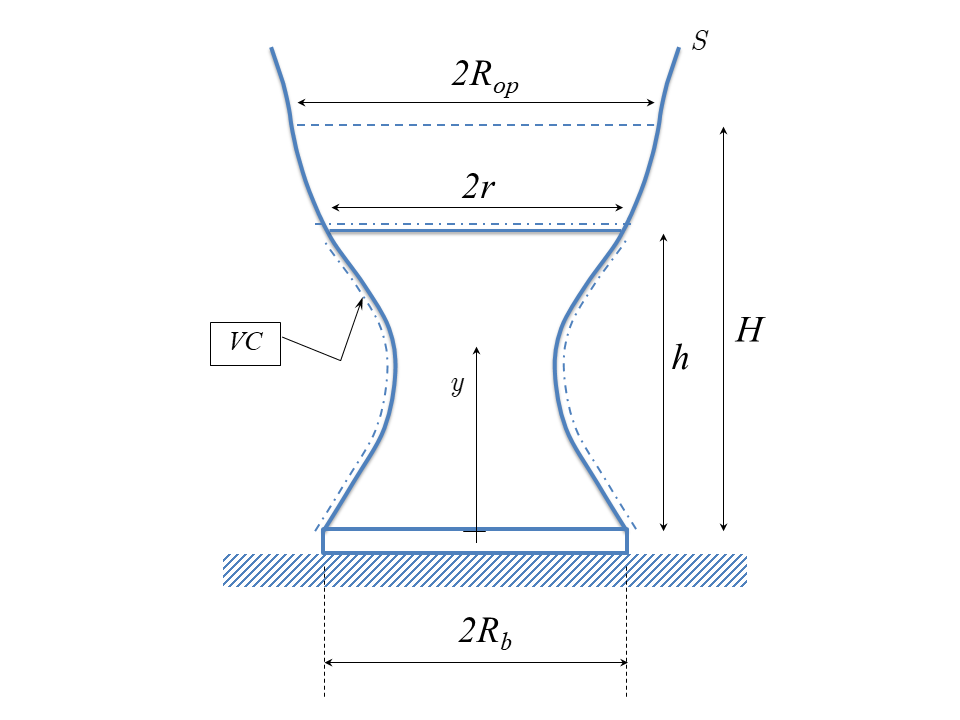}
   \caption{Typical portable liquid container device.}
   \label{fig:copo}
\end{figure}

To state the most general version of the problem, consider a container  holding a  liquid initially not in thermal equilibrium with its surroundings, as shown in Fig \ref{fig:copo}. Assume that the vessel is a body of revolution, generated by the rotation of a curve $S$, differentiable of class $C^1$, around the vertical axis, $y$. Such a  geometry  describes all commercially available containers, except for  some non-axisymmetric non-returnable beer bottles. However, $S$ is not entirely unrestricted: it must contain exactly one opening and one impermeable bottom.
 
Therefore, let $r=r(h)$ be a continuous function, differentiable of class $C^1$, describing the shape of the glass, where $h=h(t)$ is the time dependent height of liquid inside the glass. By  the inverse function theorem, $r=r(h)$ can also be written as $h=h(r)$. Both forms are used here according to convenience. 
Thus, the domain of the problem is  $0 \leq h(r) \leq H$ in the second notation or  $R_b \leq r(h) \leq R_{op}$ in the first, 
where $r(0)=R_b$ is the  radius of the  base and $r(H)=R_{op}$ is the radius of  the opening.  For obvious reasons, $r=0$ may only occur at $h=0$. 

At this point, some knowledge about Heat Transfer ir advisable. In [Pellegrini 2019] a thorough review of the concepts needed  is presented. As the text is written in Portuguese and may, thus, present difficulties for non native speakers, the textbooks by [Incropera and DeWitt 2006], [Çenjel and Ghajar 2014] and [Kreith and Bohn 2011] are also highly recommended. 

That said, let CV be a variable shape control volume\footnote{A CV is  a region of the $\mathbb{R}^3$ space chosen to study  phenomena that includes mass and/or heat transfer},  encompassing  the container and the free surface of the  liquid,  as depicted in Fig.~\ref{fig:copo}. Let's  assume  the body of the vessel has negligible thermal resistance, whereas its bottom is thermally insulated. This is a reasonable approximation for  beer glasses, where the body is made of thin glass, for comfort, and the bottom  is made of   thick glass, for mechanical resistance. Furthermore,  letting the beer glass resting over an insulating surface, such as  tables,  counters and cardboard disks (the famous "wafer"), also contributes for making the bottom insulation a reasonable hypothesis 
 
Consider that the temperature of the beer is uniform in the whole domain  while it is being consumed. This holds true when the rate of temperature change over time is small, allowing the liquid to  be in  thermal equilibrium. This is generally not the case if the initial temperature difference between the liquid and the environment is very large.

Finally, assume  the liquid to be  homogeneous. This characteristic holds true for most filtered beers, that do not form accumulations at the bottom, including the  Weizenbiers. The few  craft beers  that present small amounts of yeast sediment are not considered here.

The law of conservation of energy (\cite{incropera} for example) for the  CV chosen is  
\begin{equation}  \label{eq:plt02}
q_{ac} = q_{in} - q_{out} + q_{gen}
\end{equation}
where $q=dQ/dt$ is the heat transfer rate and $Q$ is the heat. 

In Eqn. \eqref{eq:plt02}, $q_{ac}$ represents the  accumulated heat within the CV, $q_{gen}$ is the heat generated inside the CV, and $q_{in}$ and $q_{out}$ denote the heat entering and leaving the CV, respectively.  

In general, while drinking beer and most beverages, there is no internal heat generation. Possible exceptions would be liquids undergoing fermentation. Thus, in our case, $q_{gen}=0$. Assuming that the  environment is the only external source of heat, Eqn. \eqref{eq:plt02} yields
\begin{equation}  \label{eq:plt2}
q_{ac} =  q_{in}
\end{equation}
where $ q_{in}$ enters through the body of the glass and the foam at the opening. Physically, Eqn. \ref{eq:plt2} states the obvious: the heat entering the system is accumulated and makes the liquid increase its temperature. 

A very useful analogy between heat transfer and electricity transfer is the concept of thermal resistance. In Ohms' law, $V=Ri$,  $V$ is the electric potential difference which moves electricity through the system at a flow rate $i$ against an resistance $R$. It can be shown that for simple geometries, the thermal potential difference $\Delta T$, moves  heat  against a thermal resistance $R_H$, created by all material media at a flow rate $q$. Thermal resistances can be of three types: conductive, convective and radiative, each corresponding to the heat transfer mode associated. The analogy goes as far as allowing resistances to be associated in series and in parallel.

With this concept in mind, Eqn.~\eqref{eq:plt2} can be rewritten as
\begin{equation}  \label{eq:plt3}
q_{ac} = \frac{T - T_\infty}{ R_\text{side}} +  \frac{T - T_\infty}{R_\text{FS}}
\end{equation}
where $T=T(t)$ is the spatially uniform temperature of the beer, $T_\infty$ is the ambient temperature, $R_\text{side}$ and $R_\text{FS}$ are  the thermal resistances  of the glass and  of the free surface of the liquid, respectively. 

Both resistances are composed by two other resistances associated in series: a conductive resistance  through the material and a convective resistance on  the external surfaces. However, to simplify the problem, the thermal conductivity of the side glass and the foam are neglected\footnote{Both depend on the shape of the surface, which has not yet been determined, rendering the solution iterative. }. Regarding the glass conductivity, our assumption may not be completely realistic but nevertheless represents the most critical situation, i.e., a very thin glass wall. In respect to the foam, according to [Headlee, on the internet], it decays exponentially with time, with a half time of 57 seconds. This justifies, at least in part, the second assumption\footnote{The absence of foam is considered by experts a tasting heresy}.

Equation \eqref{eq:plt3} may then be  rewritten as
\begin{equation}  \label{eq:TRC1}
	q_{ac} =  \frac{T - T_\infty} {R_\text{side}^\text{cv}}
	+ 	\frac{T - T_\infty} {R_\text{FS}^\text{cv}}
\end{equation}
In words, the preceding equations tels that the accumulated heat enters via convection heat transfer through the side of the container and the free surface of the liquid. 

From the definitions of specific heat, $c_p$, and density, $\rho$, for a volume $V$ of an homogeneous liquid it follows that $q_{ac} = \rho V c_p ({dT}/{dt})$. Thence,
\begin{equation}  \label{eq:TRC2}
\rho V c_{p} \frac{dT}{dt} =  \frac{T - T_\infty} {R_\text{side}^\text{cv}} 
+
\frac{T - T_\infty} {R_\text{FS}^\text{cv}} 
\end{equation}

According to [Incropera, 2006] 
\begin{equation}
	R^\text{cv} = 1/(h_\text{cv}A) 
\end{equation}
where $h^\text{cv}$ is the convective heat transfer coefficient in Newton's law of cooling, 
\begin{equation}
	q=h^\text{cv}A\Delta T 
\end{equation}
and $A$ is the heat transfer area. 
Equation \eqref{eq:TRC2} then finally yields 
\begin{equation}  \label{eq:TRC3}
\rho V c_{p} \frac{dT}{dt} =  h_\text{cv} A_\text{tot}(T - T_\infty)
\end{equation}
where $A_\text{tot}$ is the total heat transfer  area, i.e., the side plus the free surface area of the glass.

Under  the hypothesis used, Eqn. \eqref{eq:TRC3} shows that the temperature variation of the liquid depends  only on the  heat transfer through the exposed area of the glass and the foam and, thence that the shape of the glass is the only element responsible for  the temperature variation.

Before solving Eqn. \eqref{eq:TRC3}, it is interesting to mention  that it shows\footnote{I always tell my students that ``equations talk'' to those who can listen. You may quote me.} a number of strategies for minimizing the rate of temperature change, $dT/dt$:
\begin{enumerate}
\item Reduce $(T - T_\infty)$ by keeping the glass in a cool environment and avoiding radiant heat from the sun\footnote{Which was not considered in the present mathematical modelling and would represent an extra term in Eqn. \eqref{eq:plt3}.};
\item Increase the sum of resistances in the denominator of Eqn. \eqref{eq:TRC3} by keeping a tick, generous foam over the beer;
\item Increase the  conductivity resistance of the sides of the the vessel, by substituting the glass (the material) by a more insulating material, such as thick ceramic\footnote{This strategy is often used  on large mugs, alongside with a handle to keep the consumer's hand out of contact with the side of the glass, guaranteeing no other external heat source than the environment. Unfortunately, it makes lip contact uncomfortable, some say.};
\item Keep the glass  away from drafts, avoiding the onset of forced convection, which is far more efficient than natural convection in transferring heat.
\end{enumerate}

The considerations above  show why the beach is the most challenging environment for beer drinking: the air temperature is high,  the wind is persistent, the sun shines, and ceramic mugs are rather ridiculous. 

Rearranging Eqn. \eqref{eq:TRC3} then yields
\begin{equation}  \label{eq:TRC4}
 \frac{dT}{dt} = \frac{ h_\text{cv}}{\rho c_{p}} \left(\frac{A_\text{tot}}{V}\right)(T - T_\infty)
\end{equation}


Equation \eqref{eq:TRC4} is the ODE that governs the problem. It can be   solved for a known geometry, because  it is of the separable type. Of course, the integral involved may not have an analytical solution, but in principle it is not a challenging mathematical model. However, this would be the direct problem which is not the goal here.

For any fixed instant of time $t$ and any given value of $h_\text{cv}/ (\rho c_{p})$, Eqn. \eqref{eq:TRC4} shows that $dT/dt$ gets smaller as  $ {A_\text{tot}}/{V}$ is reduced. Therefore, the problem of minimizing the heat transfer  reduces to 
\begin{equation}
	\text{minimize}\left(\frac{A_\text{tot}}{V}\right),\quad t\in \mathbb{R}^+
\end{equation}

This is the subject of the next section.

\section{Solution}
\begin{figure} [htb]
   \centering 
   \includegraphics[trim={10mm 0 20mm 0},clip,width=0.5\columnwidth] {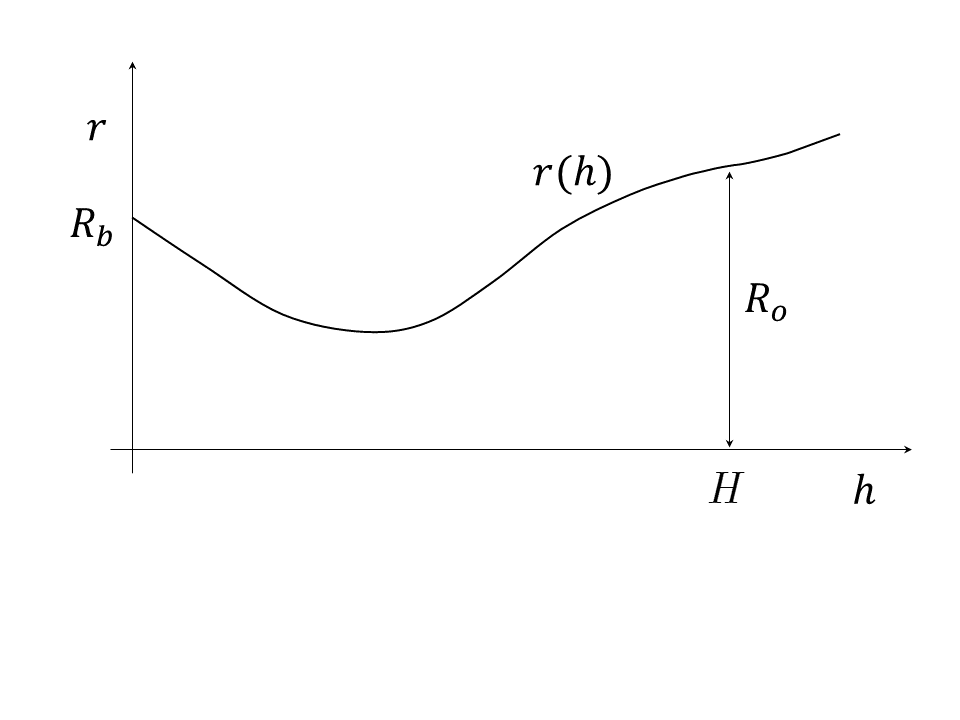}
   \caption{Glasses' generating curve}
   \label{fig:copo1}
\end{figure}

Consider Fig. \ref{fig:copo1}, where the glass and its generating curve, here identified as $r=r(h)$, are depicted

The  minimum for the surface-to-volume ratio follows from 
\begin{equation}  \label{eq:minimo1}
\frac{d}{dh} \left(\frac{A_\text{tot}}{V}\right) = 0
\end{equation}

Here, the height of the liquid, $h$, is used as the independent variable as  the beer is consumed. Differentiation of Eqn. \eqref{eq:minimo1} yields
\begin{equation}  \label{eq:minimo1.5}
\frac{V dA_\text{tot}/dh - A_\text{tot} dV/dh} {V^2}=0
\end{equation}

Simplifying  and integrating results in
\begin{equation}  \label{eq:minimo2}
\int\frac{dA_\text{tot}}{A_\text{tot}} =\int\frac{dV}{V}
\end{equation}
for  $V \neq 0$ and $A_\text{tot}\neq 0$. 
The result is $\ln({A_\text{tot}}) =\ln( V)+ C$ and, thus,
\begin{equation}  \label{eq:minimo3}
A_\text{tot}=C_1 V
\end{equation}
where $C_1$ does not depend on $h$ and have dimension $L^{-1}$.

From Calculus,  bodies of revolution have side areas and the volumes  given by
\begin{equation}  \label{eq:area}
A_\text{lat} = 2\pi \int_0^h{r\sqrt{1+r'^2}\,dh} 
\end{equation}
and
\begin{equation}  \label{eq:volume}
V = \pi  \int_0^h r^2\,dh
\end{equation}
where the total area of heat transfer is $A_\text{tot}=A_\text{lat}+\pi r^2$. Equation \eqref{eq:minimo3} then becomes
\begin{equation}  \label{eq:minimo4}
\pi r^2 + 2\pi \int_0^h{r\sqrt{1+r'^2}\,dh} = C_1 \pi  \int_0^h r^2\,dh
\end{equation}

Differentiating in relation to $h$ results in\footnote{Note that the derivative of an integral is the integrand itself as long as the upper integration limit coincides with the differentiated variable, as is the case here.}
\begin{equation} \label{eq:minimo4.3}
	2\pi r r' + 2\pi r \sqrt{1+r'^2} = C_1 \pi r^2
\end{equation}

Simplifying and rearranging,
\begin{equation} \label{eq:minimum5}
	\sqrt{1+r'^2} =C_1 r/2 - r'
\end{equation}
for $r \neq 0$, except possibly at $h=0$. 
Squaring both sides and simplifying,
\begin{equation} \label{eq:edo1}
	1 = (C_1/2)^2 r^2 - C_1rr'
\end{equation}

This is a separable ODE. Therefore,
\begin{equation} \label{eq:edo2}
	\int_0^h{dh} = \int_{R_b}^r{ \frac{C_1r}{(C_1/2)^2 r^2 - 1}dr}
\end{equation}
which can be integrated by substitution of $\eta = (C_1/2)^2 r^2 - 1$. The result is
\begin{equation} \label{eq:h1}
	h = \frac{2}{C_1}\ln \left[{(C_1/2)^2 r^2 - 1}\right] - \frac{2}{C_1}\ln \left[{(C_1/2)^2 R_b^2 - 1}\right]
\end{equation}

Rearranging to make $r$ explicit gives
\begin{equation} \label{eq:h}
	r = \pm\frac{1}{C_1}\sqrt{4 + (C_1^2 R_b^2 - 4)e^{C_1 h/2}}
\end{equation}
which solves the problem.

To obtain a equation for the volume of the glass,  Eqn. \eqref{eq:h} is substituted into Eqn.  \eqref{eq:volume} yielding, upon integration, 
\begin{equation} \label{eq:volume3}
	V = \frac{\pi}{C_1^3}\left[4C_1 h + 2(C_1^2 R_b^2-4) \left(e^{C_1 h/2}-1\right)\right]
\end{equation}

This equation may now  substituted into Eq. \eqref{eq:minimo3} giving
\begin{equation} \label{eq:area3}
	A_\text{tot} = \frac{\pi}{C_1^2}\left(4C_1 h + (C_1^2 R_b^2-4) \left(e^{C_1 h/2}-1\right)\right)
\end{equation}
and 
\begin{equation} \label{eq:area4}
	A_\text{lat} = \frac{\pi}{C_1^2}\left(4C_1 h + (C_1^2 R_b^2-4) \left(e^{C_1 h/2}-1\right)\right) - \pi r^2
\end{equation}

The preceding  equations express the dependence of the volume and the wet surface of the glass with the height of the liquid {\sl as it is consumed}. The total volume is obtained simply by putting $h=H$ into Eq.  \eqref{eq:volume3}. To obtain  the total surface of the glass, $r$ must be substituted by $R_{op}$ in Eq. \eqref{eq:area4} and the term $2\pi R_\text{op} h_\text{f}$ must be added, with $ h_\text{f}$ being the vertical distance from the surface of the liquid to the edge of the glass, a space reserved to the necessary foam. The approximate result is then
\begin{equation}
		A_\text{lat} \approx \frac{\pi}{C_1^2}\left(4C_1 h + (C_1^2 R_b^2-4) \left(e^{C_1 h/2}-1\right)\right) - \pi R_b^2 + 2\pi R_\text{op} h_\text{f}
\end{equation}

This result is not exact, because it assumes that the upper part of the glass is cylindrical, which is seldom the case.

\section{Discussion}
It is easy to see that the function  $r(h)$, that minimizes the heat transfer, increases  from $r=R_b$ to $r=R_{op}$ for $C_1 > 0$. In other words, that  the glass  have a small base and a large opening as expected, provided that $C_1 > 0$. To show that  $r(h)$ grows monotonically, first, Eq. \eqref{eq:h} is  rewritten as 
\begin{equation} \label{eq:h2}
	h = \frac{2}{C_1} \ln \left(\frac{ C_1^2 r^2 - 4} {C_1^2 R_b^2 - 4}\right)
\end{equation} 
which implies that $C_1r\neq 2$. But if the condition for the existence of a maximum,  $r'=0$, is substituted  into Eqn. \eqref{eq:edo1} it follows  that $C_1 r=2$. Thence, by contradiction, $r(h)$ must grow monotonically. 

\begin{figure}[htb]
	\centering
	\includegraphics[width=0.7\linewidth]{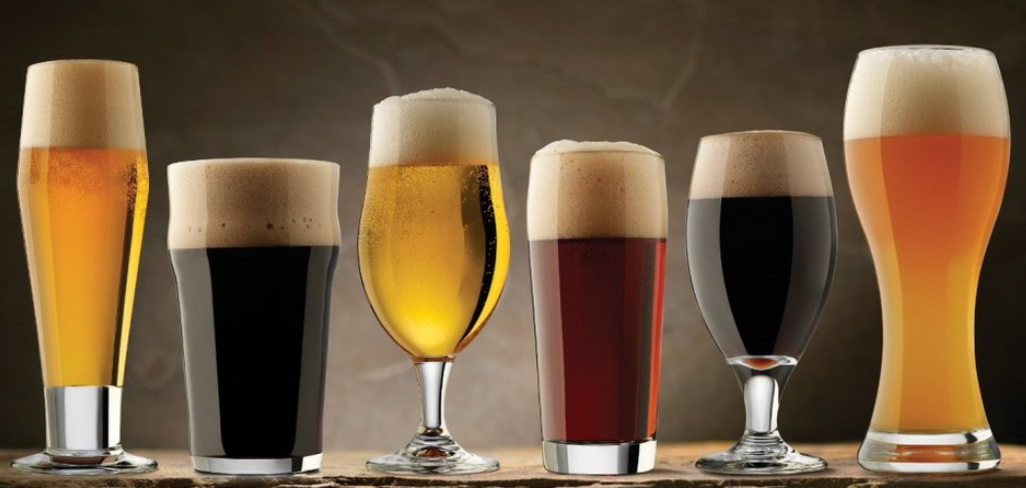}
	\caption{Some types of glasses commercially available. Each is considered appropriate to a particular kind of beer. Only the second and the fourth from the left vary (approximately) monotonically. Source: https://www.crateandbarrel.com/ideas-and-advice/types-of-beer-glasses.}
	\label{fig:types}
\end{figure}

Such behavior excludes many glass types, as illustrated by Fig. \ref{fig:types}. It will soon be demonstrated that this restriction  generates  a family of glasses. But if the negative sign before the square root on Eqn. \eqref{eq:h} is considered, a family of glasses described by a monotonically decreasing function is generated. This possibility is currently under investigation,  and will not be pursued any further here. For the moment, $C_1>0$ will be treated as a working hypothesis and its implication in  Eqn. \eqref{eq:h2} is 
\begin{equation} \label{eq:restricao2a1}
	\frac{ C_1^2 r^2 - 4}{ C_1^2 R_b^2 - 4}>0
\end{equation}	

This inequality also has more than one  implication, but just simplest one will be discussed here, i.e.,  
\begin{equation} \label{eq:restricao2a}
	C_1 r \neq 2
\end{equation}
which is a negative relation that will soon prove to be very important. Indeed, before  Eqn. \eqref{eq:h} is sent to the factory, and the optimized glasses start to be mass produced, it is necessary to establish adequate values for $C_1$, limited by Eqn. \ref{eq:restricao2a}.



In Eqn. \eqref{eq:h2}, as  $C_1 R_b \rightarrow 2$ from the right for fixed values of $r$, then $h \rightarrow \infty$ for any finite $C_1$. In words, as $C_1 R_b \rightarrow 2$, the optimal solution is only achievable for very tall glasses. This should not be an issue, as one can simply chose  a large value of $C_1 R_b$. However, as $C_1 R$ multiplies $\exp(C_1h/2)$, large values of $C_1 R_b$  results in very large values of $r$. Thus as $C_1 R_b \gg 2$  the optimal solution can  only be obtained for very large openings. Both situations suggest that there exists a suitable range of values for $C_1 R_b$ that results in practically viable glass shapes. 

To obtain plausible values of $C_1$, Eqn. \eqref{eq:h2} is initially  rewritten as
\begin{equation} \label{eq:h3}
	e^{(C_1h/2)} = \frac{ C_1^2 r^2 - 4} {C_1^2 R_b^2 - 4}
\end{equation}

Then, based on practical considerations, values of $R_{op}$  and $\lambda = R_{op} / R_b$ are chosen. Substituting   $r=R_{op} = \lambda R_b$ and $h=H$ into Eqn.~\eqref{eq:h3} then yields
\begin{equation} \label{eq:h4}
	e^{(C_1H/2)} = \frac{ (\lambda C_1 R_b)^2 - 4}{(C_1 R_b)^2 - 4}
\end{equation}

Supposing that $ {C_1 R_b} \approx 2 $ (to avoid  tall glasses) and substituting into  Eqn.~\eqref{eq:h4} gives
\begin{equation} \label{eq:h5}
	e^{(C_1H/2)} = \frac{ 4(\lambda^2 -1)}{ \varepsilon}
\end{equation}
where $ \varepsilon = {C_1^2 R_b^2 - 4}  \approx 0$ is a small number. 
Equation \eqref{eq:h4} then becomes 
\begin{equation} \label{eq:eps}
	{ \varepsilon} \approx { 4(\lambda^2 -1)}e^{-C_1H/2}
\end{equation}

Finally, the definition off $\varepsilon$ can be inverted to yield
\begin{equation} \label{eq:eps2}
	C_1R_b = \sqrt{ \varepsilon + 4}
\end{equation}

Therefore, Eqs. \eqref{eq:eps} and \eqref{eq:eps2} can be used to estimate $C_1$ in the following way:
\smallskip
\begin{enumerate}
	\item Chose adequate values for $R_{op} $ and $R_b $;
	\item Obtain a first approximation for $C_1$  considering $\varepsilon=0$ in Eq. \eqref{eq:eps2}. This value is not final because it results in division by zero in Eqn. \ref{eq:h3}, once  $C_1R_b=2$;
	\item Substitute the value of $\varepsilon$ into Eqn. \eqref{eq:eps}  and  recalculate $\varepsilon$;
	\item Substitute this value  of $\varepsilon$  into Eqn. \eqref{eq:eps2}.
\end{enumerate}

As  $C_1$ is only as a shape parameter in the solution, there is no need for  further iterations or for establishing a convergence criterion. The shape of the glass can be obtained from Eqn. \eqref{eq:h}.

For some obscure reason, in [Pellegrini, 2019] the author let  $C_1$ vary around the value obtained in step (2) instead of following steps (3) and (4). The procedure yielded a few feasible solutions, but most calculations returned ridiculously large glasses, with volumes of the order of 100 liters. 

Indeed, proposing and following the above steps is one of the contributions of the present work.

Therefore, a few typical categories of commercial glasses were chosen to test the methodology. According to the step (1), a value for $\lambda$ was chosen and lower limit to the value for $R_{b}$ was imposed. The value of $C_1$ was then calculated according to steps (2)--(4) yielding an optimized glass. If the volume of this glass differed from the real glass' volume by more than a small prescribed   value (1 ml), $R_{b}$ was then incremented by a small quantity (0.1 mm) and the calculation was repeated. Therefore, the process used as inputs the height of the glass and its relation between the base and the opening diameter and returned  the base radius of the optimized glass with almost the same volume of the real one. 
The procedure was implemented in an extremely simple computational code written in MatLaB\textsuperscript{\textregistered} student version. 
The results are presented and discussed in the next section.s.

\section{Results}
\begin{figure} [htb]
   \centering 
   \includegraphics [width=0.5\columnwidth] {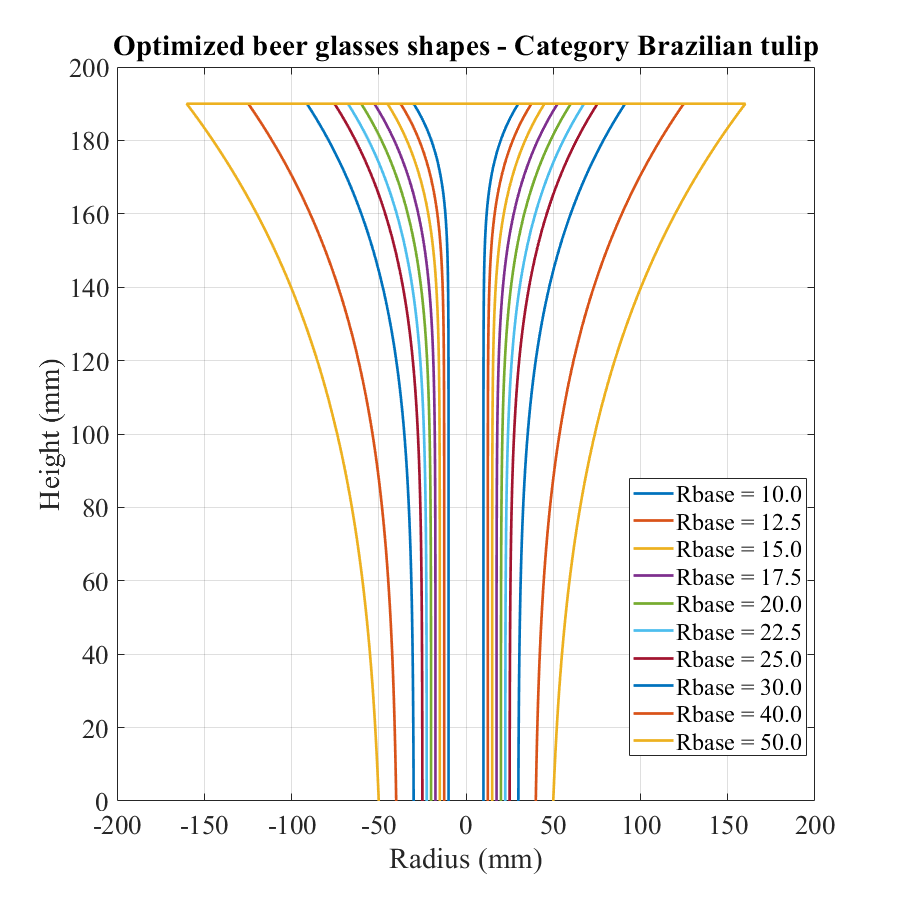}
   \caption{Optimized shapes for $H= 190$ mm and $\lambda=3$ for varying values of $R_b$.}
   \label{fig:tulips}
\end{figure}

Figure \ref{fig:tulips} shows a typical result of a run of the code $H$ and $\lambda$  keep constant and $R_b$ increasing from . It i easy to see that large glasses are not simply a magnified version of the small. Compare, for example, glasses with $R_{b}=10$ mm and $R_{b}=50$ mm -- the outermost and innermost lines respectively. Whereas the former is curved all the way from base to opening, the later is almost cylindrical up to 1500 mm of height. 

\begin{figure} [htb]
	\centering 
	\includegraphics [width=0.31\columnwidth] {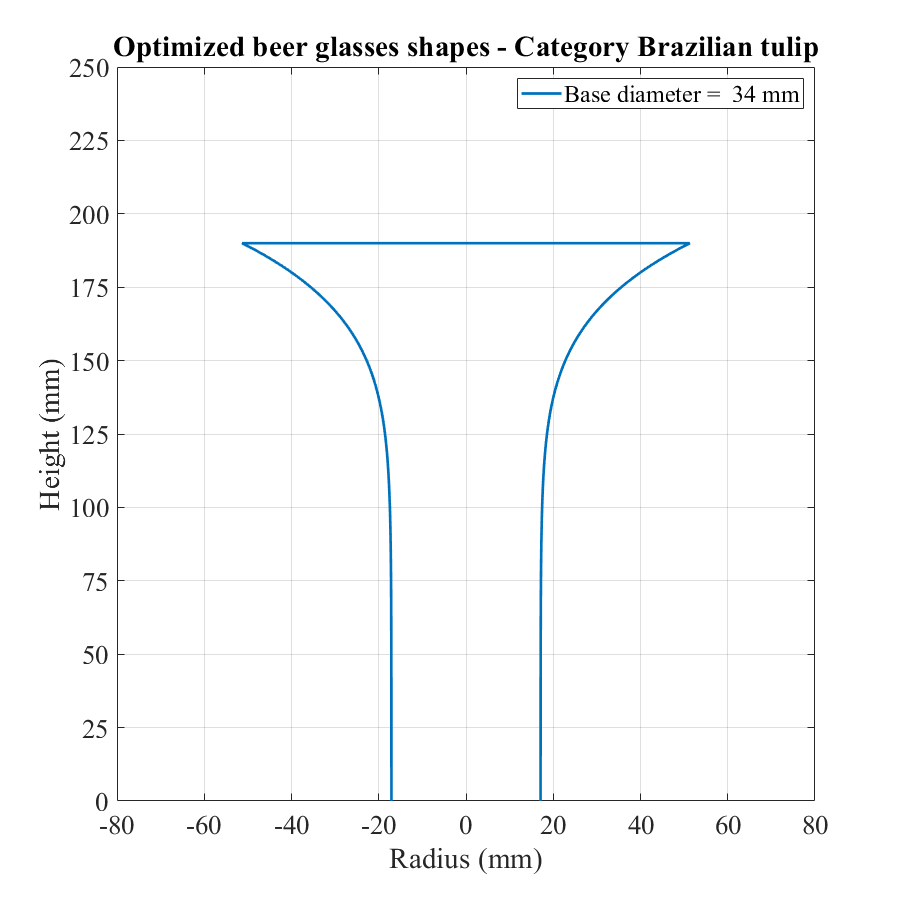}
	\includegraphics [width=0.31\columnwidth] {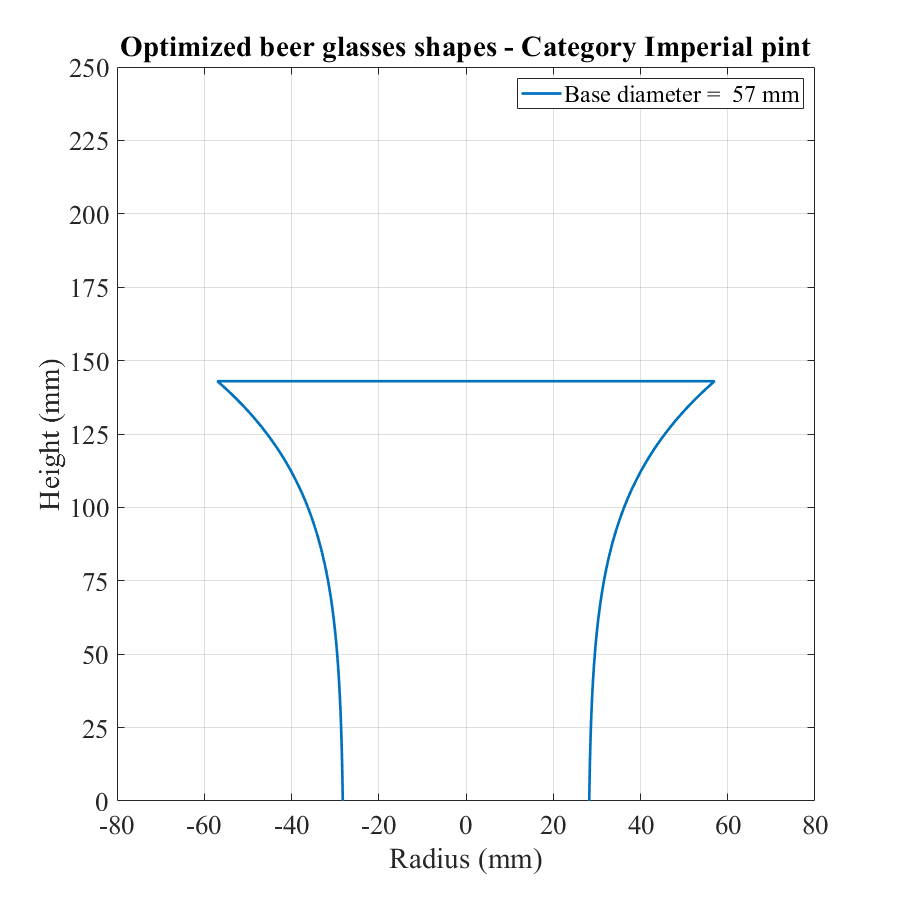}	\includegraphics [width=0.31\columnwidth] {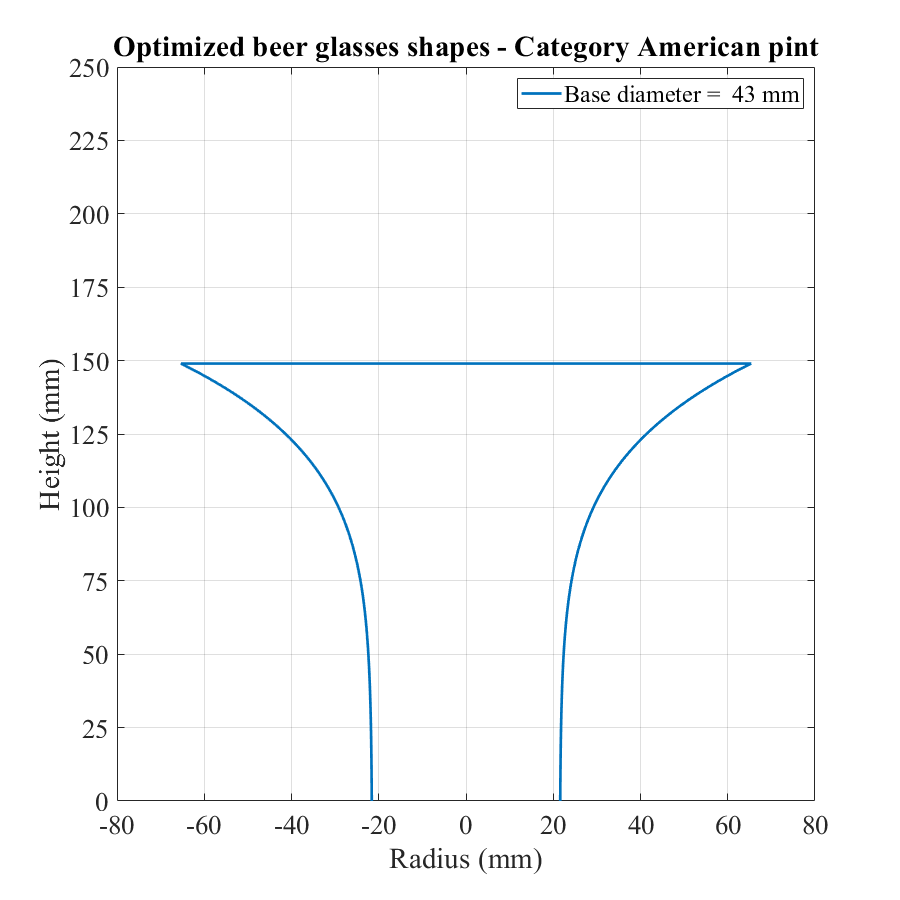}
	\caption{Optimum glass of the categories Brazilian tulip (left)  Imperial pint (center) and American pint (right).}
	\label{fig:best1}
\end{figure}

\begin{figure} [htb]
	\centering 
	\includegraphics [width=0.31\columnwidth] {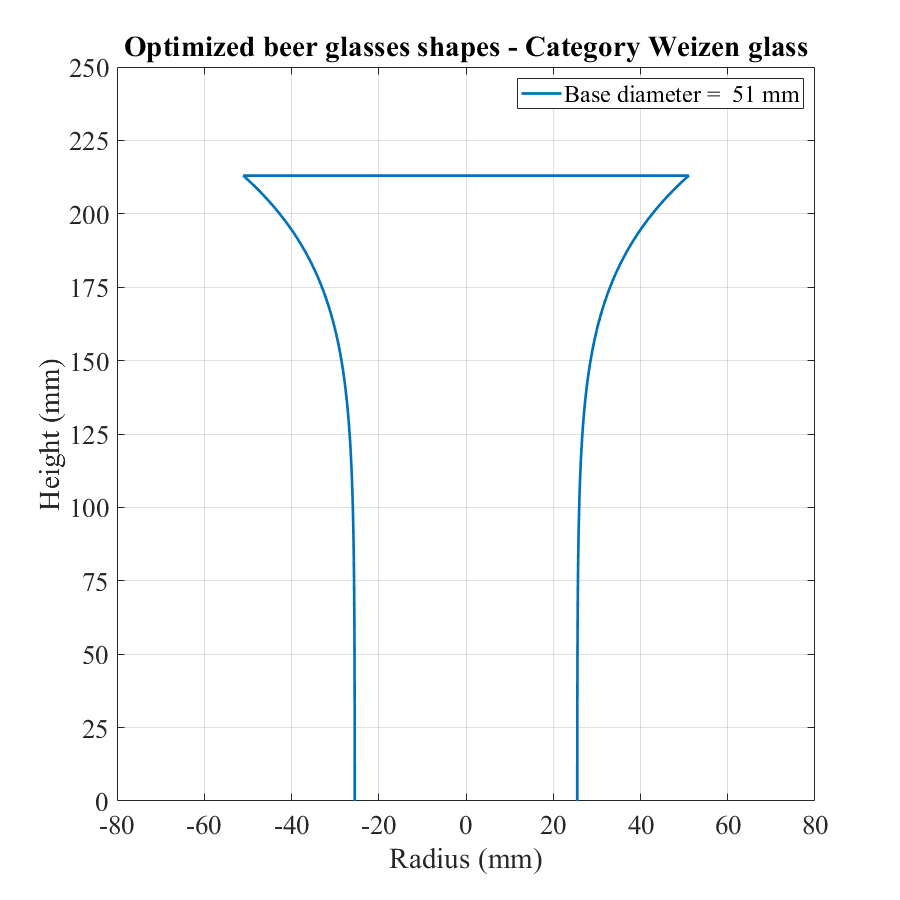}
	\includegraphics [width=0.31\columnwidth] {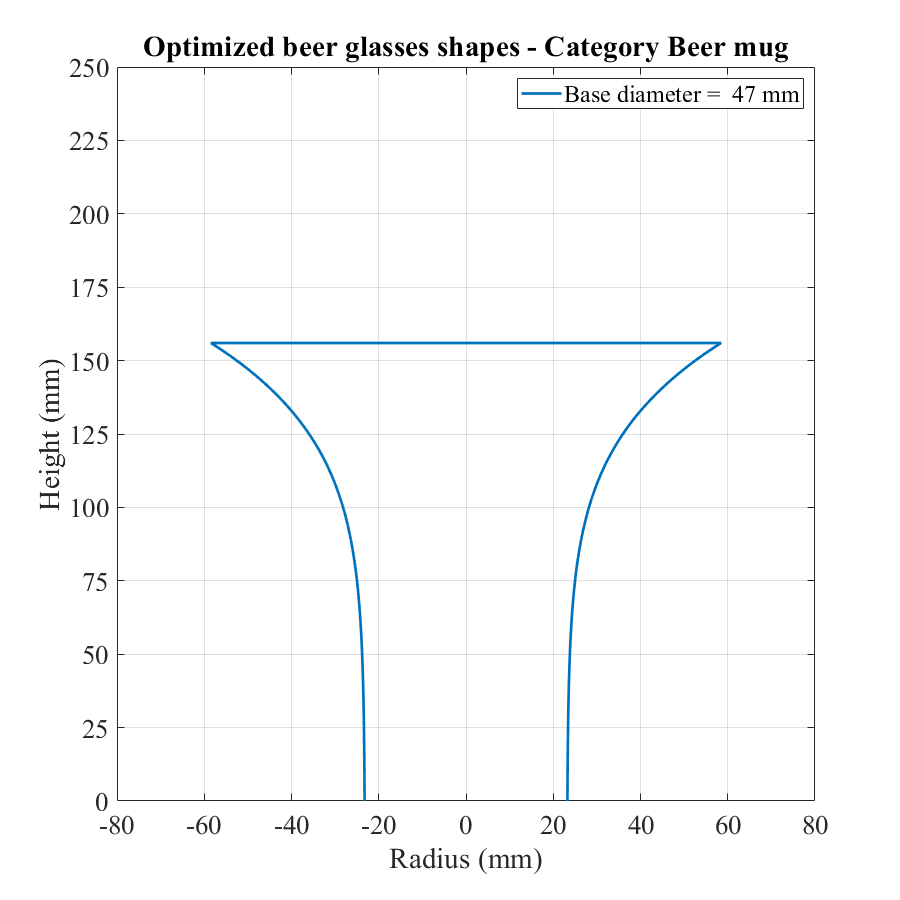}	
	\caption{Optimum glasses of the categories Weinzen glass (left), and Beer mug (right).} 
	\label{fig:best2}
\end{figure}

\begin{figure} [htb]
	\centering 
	\includegraphics [width=0.31\columnwidth] {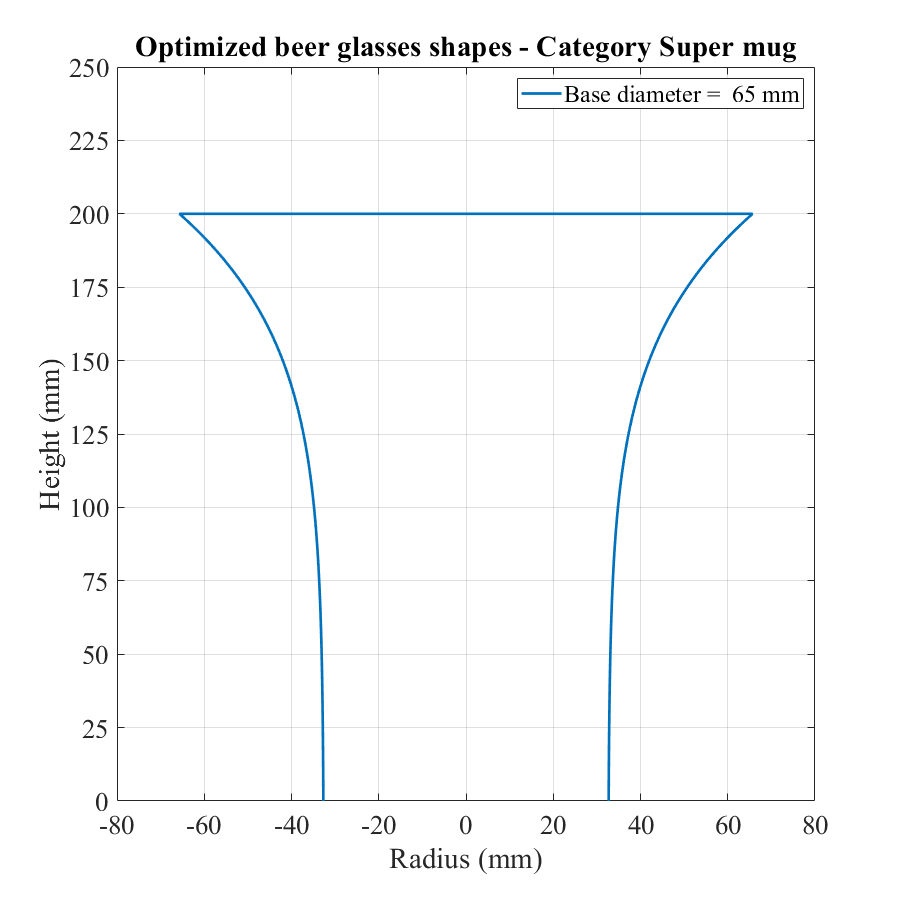}	
	\includegraphics [width=0.31\columnwidth] {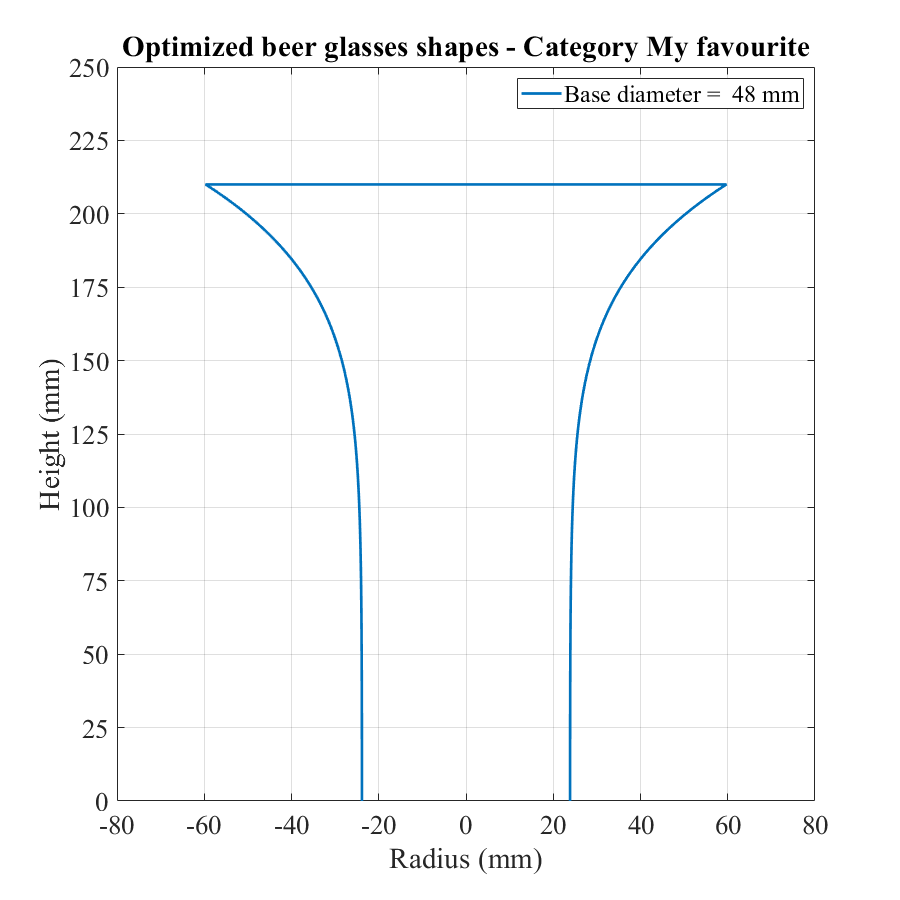}
	\caption{Optimum glasses of the categories Super mug (left) and My favorite (right).} 
	\label{fig:best3}
\end{figure}

The proposed methodology was applied to five traditional  categories of glasses: the Brazilian tulip, the Imperial pint, the American pint,   the Weizen glass, and the Beer Mug\footnote{Their values of height, volume and opening radius are available on the internet: https://www.dimensions.com/collection/beer-glasses.}. Their dimesions are shown in Table \ref{tab:dimensions1}. Two non-standard  categories  were also introduced: the Super mug and the My favorite,   reflecting the author's preferences. The code was then run for each category and the results are depicted in  Figs.~\ref{fig:best1}~--~\ref{fig:best3}. Numerical results appear in Table \ref{tab:dimensions2}. 
\begin{table}[htb]
	\centering
		\caption{Typical dimensions of the glasses' categories considered.}
	\begin{tabular}{|l|c|c|c|c|}
		\toprule
		{\bf Category} &{$H$ (mm)} & {$D_{op}$ (mm)} & {$V$ (ml)} \\
		\midrule
		Brazilian tulip & 190 &  105 & 300 \\
		Imperial Pint 	& 143 &  120 & 568 \\
		American pint 	& 149 &  100 & 473 \\
		Weizen glass 	& 213 &  79	 & 591 \\
		Beer mug		& 156 &  89	 & 473 \\
		Super mug 		& 200 &  120 & 1000\\
		My favorite 	& 210 &  100 & 600 \\
		\bottomrule
	\end{tabular}
	\label{tab:dimensions1}
\end{table}

\begin{table}[htb]
	\centering
	\caption{Dimensions of the optimized glasses.}
	\begin{tabular}{|l|c|c|c|c|c|c|}
		\toprule
		{\bf Category} & $H$ (mm) & $\lambda$ &  $D_b$ (mm) & $V$ (ml) & $\Delta V$ (ml) & $C_1$ (1/m)\\
		\midrule
		Brazilian tulip & 190 & 3.0 & 34 & 300 & 0.2 & 117.0\\
		Imperial Pint 	& 143 & 2.0 & 57 & 567 & 0.8 & 71.5 \\
		American pint 	& 149 & 3.0 & 43 & 474 & 0.7 & 93.0 \\
		Weizen glass 	& 213 & 2.0 & 51 & 591 & 0.3 & 78.5 \\
		Beer mug		& 156 & 2.5 & 47 & 473 & 0.2 & 86.3\\
		Super mug 		& 200 & 2.0 & 66 & 1000& 0.3 & 61.4 \\
		My favorite 	& 210 & 2.5 & 48 & 599 & 0.8 & 83.9 \\
		\bottomrule
	\end{tabular}
	\label{tab:dimensions2}
\end{table}


Some details  are worth emphasizing at this point. First, all solutions are very similar, despite the glass category being considered. This is a direct consequence of the fact that the solution is unique, i.e., that Eq. \eqref{eq:h} has one single root in the positive real  numbers for $C_1>0$. Therefore, the lateral curves are qualitatively similar, being  only quantitatively modified by $H$, $R_b$ and $\lambda$. 

Other important aspect is that each optimized glass represents the ideal solution for a specific configuration, i.e., its chosen  values of $H$, $R_b$ and $\lambda$. There is no way to verify if, for example, the best Weizen glass with $R_b=25$ mm   is better than the best Weizen glass with $R_b=30$ mm. So far, our methodology only guarantees that each one is the best  {in its own category}, for a given configuration. The same is valid when comparing glasses of  different categories, as for example, the best Brazilian tulip and the best Imperial pint. This shortcoming is being addressed at the moment and will, with some luck,  be published briefly.

Finally, taken as a group, the seven configurations proposed  cover a wide interval of  volumetric capacities, ranging from the relatively small  Brazilian tulip (300 ml), conceived to drink  the mass-produced Brazilian Pilsens in hot weather, to  the super mug, a configuration proposed by the author to meet the expectations of the bold, with one liter capacity. 

As a side note for those unfamiliar with Brazilian's drinking habits, beer is sold in a multitude of glass bottles and cans. The most common bottle has a volume of 600 ml (thus the volume of the My Favorite category), but bottles bearing 355, 330, 550 and 1000 ml are also common. Beer cans are available in even more capacity values, with typical examples being 269, 300, 330,  355, 473, 500 ml. I am sure that as soon as this research is publish, another capacity will have already  been invented.

Regarding the types of glass used here, the Brazilian tulip is a shape  widely used in bars, restaurants and, quite inappropriately, on our beaches. It is a rather small  glass, bearing  only 300 ml capacity. Many  believe it to be the natural, fancy substitute of the most widely used of all glasses in Brazil, the ``American'' glass. 

\begin{figure}[htb]
	\centering
	\includegraphics[width=0.20\linewidth]{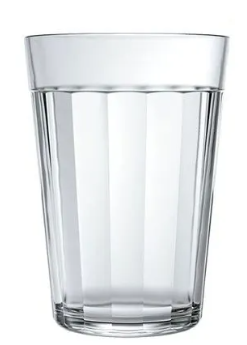}
	\caption{The infamous Brazilian's Nadir Figueiredo glass.}
	\label{fig:NF}
\end{figure}

The American glass, also known as the ``Nadir Figueiredo\textsuperscript{\textregistered}  glass'', due to the family name of  the first industry to manufacture it, contains a rather low volume (190 ml) and is exceptionally ugly (Fig. \ref{fig:NF}). Still, it is the most widely used beer glass  in Brazil, counting more than 6 billion units manufactured since its introduction in the 40's. It is considered an icon in beer drinking and costs less than a quarter dollar\footnote{Yes, the low price is also important when when choosing the glass to a particular event. Depending upon the moods of the gathering, people break as many as 20\% of all  glasses used.}. It maintains the beer cold through the most primitive of the processes: due to its low capacity, the beer is consumed so quickly it has no time to get warm\footnote{And no, you don´t know somebody that drinks {\sl that} slow.}.

\section{Conclusions}
In this work, a method was proposed to optimize the shape of beer glasses, with the goal of minimizing heat transfer and maintaining a low liquid temperature for an extended period during consumption. The analysis resulted in a family of shapes that can be easily manufactured using traditional methods and are suitable for everyday use.

Throughout the analysis several hypothesis were made. The glass was modeled as a body of revolution generated by the rotation of a  continuous, class $C^1$, monotonically increasing curve  around the vertical axis. Additionally,   thermal resistance in the glass body  was neglected, and the bottom was assumed to be  insulated. The liquid's temperature was considered spatially uniform, and  the liquid itself was treated as homogeneous, with the thermal resistance of the foam disregarded.  Finally, neither radiative heat transfer nor conduction due to hand contact with the glass was considered. 

While these hypotheses may seem restrictive, the analytical results obtained remain of both didactic and practical interest. A more complete formulation, incorporating the effects previously disregarded, can be effectively addressed through numerical methods. However, since the primary focus of this investigation is didactic, analytical solutions are preferable. Closed-form analytical solutions are often favored in physics, even when based on simplified analyses, as they offer a clear and explicit representation of the influence of all parameters involved

Moreover, an analytical solution typically provides a general conclusion about the problem rather than focusing on a specific case study. It also clarifies the conditions under which the results are valid. While these points may seem obvious, they are particularly relevant in an era marked by the widespread and sometimes careless use of computer simulations and artificial intelligence.

The problem addressed here is far from fully solved, and further studies are needed. Future investigations should consider heat transfer through the base of the glass, account for radiative heat transfer, and include the presence of foam. Additionally, developing a criterion to obtain the global optimum for glasses across various categories is an ongoing line of research.

In conclusion, this paper demonstrated how basic concepts of heat transfer and extreme values of functions can be applied to a relevant everyday topic -- best practices in beer drinking. Despite the light tone, the analysis was presented with appropriate mathematical rigor. 

The primary goal of this paper is to enhance the interest of students in exact sciences, particularly in Physics and Mathematics. Additionally, a secondary yet crucial application of these findings is to educate future generations of beer enthusiasts and to safeguard the quality of our beers



\end{document}